\documentclass[11pt]{article}
\usepackage{graphicx}
\usepackage{epsfig}

\newcommand{\BABARPubYear}    {03}

\newcommand{\BABARConfNumber} {002}
\newcommand{\SLACPubNumber} {9667}

\RequirePackage{xspace}
\usepackage{relsize}
\def\qqbar {\ensuremath{q\overline q}\xspace}
\def\babar{\mbox{\slshape B\kern-0.1em{\smaller A}\kern-0.1em
    B\kern-0.1em{\smaller A\kern-0.2em R}}}
\def\Bbar    {\kern 0.18em\overline{\kern -0.18em B}{}\xspace}

\def\BB      {\ensuremath{B\Bbar}\xspace} 
\def\Bz      {\ensuremath{B^0}\xspace}
\def\Bzb     {\ensuremath{\Bbar^0}\xspace}
\def\BzBzb   {\ensuremath{\Bz {\kern -0.16em \Bzb}}\xspace}
\def\Bu      {\ensuremath{B^+}\xspace}
\def\Bub     {\ensuremath{B^-}\xspace}

\def\BpBm    {\ensuremath{\Bu {\kern -0.16em \Bub}}\xspace}

\newcommand{\optbar}[1]{\shortstack{{\tiny (\rule[.4ex]{1em}{.1mm})}
  \\ [-.7ex] $#1$}}
\def\BorBbar    {\kern 0.18em\optbar{\kern -0.18em B}{}\xspace}
\def\DorDbar    {\kern 0.18em\optbar{\kern -0.18em D}{}\xspace}
\def\KorKbar    {\kern 0.18em\optbar{\kern -0.18em K}{}\xspace}
\def\CP                {\ensuremath{C\!P}\xspace}
\def\pep2{PEP-II}
\mathchardef\Upsilon="7107
\def\Y#1S{\ensuremath{\Upsilon{(#1S)}}\xspace}% no space before {...}!

\def\FourS {\Y4S}

\long\def\inst#1{\par\nobreak\kern 4pt\nobreak
    {\it #1}\par\vskip 10pt plus 3pt minus 3pt}

\begin{document}
{\pagestyle{empty}

\begin{flushright}
\babar-CONF-\BABARPubYear/\BABARConfNumber \\
%\babar-PUB-\BABARPubYear/\BABARPubNumber \\
SLAC-PUB-\SLACPubNumber \\
%hep-ex/\LANLNumber \\
March 2003 \\
\end{flushright}

%\par\vskip 5cm
\par\vskip 3cm

% Title of the paper
\begin{center}
\Large \bf 
\boldmath
Rates, Polarizations, and Asymmetries in Charmless Vector-Vector $B$ Decays
\end{center}
\bigskip

\begin{center}
\large The \babar\ Collaboration\\
\mbox{ }\\
%\today
March 13, 2003 \\
\end{center}
\bigskip 
%\bigskip

% Abstract
\begin{center}
\large \bf Abstract
\end{center}
With a sample of approximately 89 million $\BB$
pairs collected with the $\babar$ detector, 
we measure branching fractions, 
determine the degree of longitudinal 
polarization, and search for direct $\CP$ violation in
the decays $B^0\to\phi K^{*0}$ and $B^+\to\phi K^{*+}$.
We perform a search for other charmless vector-vector 
$B$ decays involving $\rho$ and $K^*(892)$ resonances
and observe the decays $B^+\to\rho^0 K^{*+}$ and
$B^+\to\rho^0 \rho^+$.
The branching fractions are measured to be
${\cal B}(\phi K^{*0})=(11.1^{+1.3}_{-1.2}\pm 1.1)\times 10^{-6}$,
${\cal B}(\phi K^{*+})=(12.1^{+2.1}_{-1.9}\pm 1.5)\times 10^{-6}$,
${\cal B}(\rho^0 K^{*+})=(7.7^{+2.1}_{-2.0}\pm 1.4)\times 10^{-6}$,
and
${\cal B}(\rho^0 \rho^{+})=(9.9^{+2.6}_{-2.5}\pm 2.5)\times 10^{-6}$.
The longitudinal polarization fractions 
are measured to be
${\Gamma_L}/{\Gamma}(\phi K^{*0})=0.65\pm 0.07\pm 0.04$ and
${\Gamma_L}/{\Gamma}(\phi K^{*+})=0.46\pm 0.12\pm 0.05$.
We measure the charge asymmetries:
${\cal A}_{\CP}(\phi K^{*0})=+0.04\pm 0.12\pm 0.02$ and
${\cal A}_{\CP}(\phi K^{*+})=+0.16\pm 0.17\pm 0.04$.

\vfill
\begin{center}
Presented at the XVII$^{th}$ Rencontres de la Vall\'ee d'Aoste, \\
3/9---3/15/2003, La Thuile, Vall\'ee d'Aoste, Italy
\end{center}

\vspace{1.0cm}
\begin{center}
{\em Stanford Linear Accelerator Center, Stanford University, 
Stanford, CA 94309} \\ \vspace{0.1cm}\hrule\vspace{0.1cm}
Work supported in part by Department of Energy contract DE-AC03-76SF00515.
\end{center}

\newpage
} % end of pagestyle{empty}

% Author list file
%\input pubboard/authors_win2003.tex
%%%%%%%%%%%%%%%%%%%%%%%%%%%%%%%%%%%%%%%%%%%%%%%%%%%%%%%%%%%%%%%%%%%%%%
\begin{center}
\small

The \babar\ Collaboration,
\bigskip

%% author list as of 01-Feb-2003 (555 authors)
%
B.~Aubert,
R.~Barate,
D.~Boutigny,
J.-M.~Gaillard,
A.~Hicheur,
Y.~Karyotakis,
J.~P.~Lees,
P.~Robbe,
V.~Tisserand,
A.~Zghiche
\inst{Laboratoire de Physique des Particules, F-74941 Annecy-le-Vieux, France }
% ...

\end{center}\newpage
%%%%%%%%%%%%%%%%%%%%%%%%%%%%%%%%%%%%%%%%%%%%%%%%%%%%%%%%%%%%%%%%%%%%%%
% The body of the paper starts here
%%%%%%%%%%%%%%%%%%%%%%%%%%%%%%%%%%%%%%%%%%%%%%%%%%%%%%%%%%%%%%%%%%%%%%
\section{INTRODUCTION}
\label{sec:Introduction}

The first evidence for charmless vector-vector $B$ decays 
was provided by the CLEO experiment with the measurement 
of one $B\to{\phi K^{*}}$ channel \cite{cleo}. 
Measurement of both $B\to{\phi K^{*}}$ charge modes was 
performed by \babar~\cite{prlphik}, which also reported 
the search for direct $\CP$ violation in these 
modes~\cite{prdacp}.
The CLEO experiment also set upper limits on the decay rates
in several other vector-vector final states of $B$ decays 
\cite{cleorhokst}.
The Belle experiment announced large signal
yield in the $B^+\to{\rho^0\rho^+}$ channel~\cite{belle}.

Recently, there has been interest in charmless 
$B$ decays because of the clean environment for the search 
for new physics. For example, new particles contributing 
to penguin diagrams, such as charged Higgs bosons or SUSY 
particles, would provide additional amplitudes with different 
phases. Charmless $B$ decays are also sensitive to the 
weak phases 
$\alpha\equiv {\rm arg}\,[\, - V^{ }_{td}V^*_{tb}\,/\,V^{ }_{ud}V^*_{ub}\,]$
and 
$\gamma\equiv{\rm arg}\,[\,-V^{ }_{ud}V^*_{ub}\,/\,V^{ }_{cd}V^*_{cb}\,]$ 
arising from the elements of the Cabibbo-Kobayashi-Maskawa (CKM)
mixing matrix~\cite{Kobayashi}.
The vector-vector charmless $B$ decays provide 
additional information about the decay dynamics and 
strong phases, which could be obtained from the analysis 
of angular distributions~\cite{bvv}.

\begin{figure}[htbp]
\begin{center}
\centerline{
\epsfig{figure=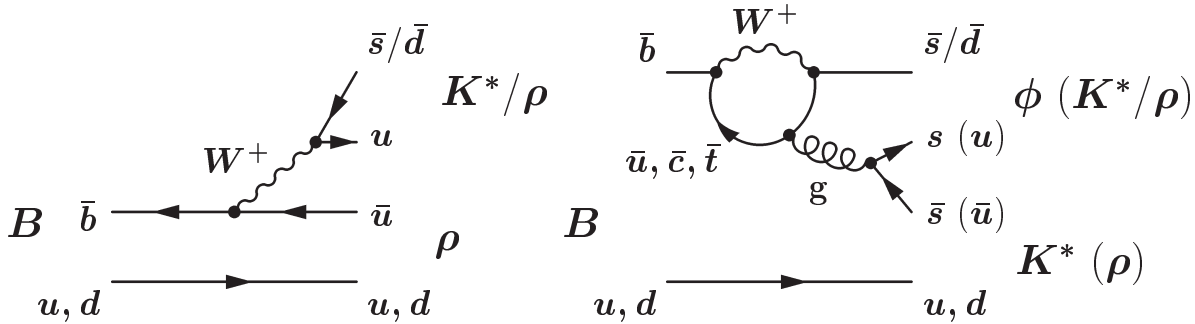,height=2.0in}
}
\caption{\sl 
Two of the dominant diagrams describing the decays 
$B\to\rho\rho$, $\rho K^{*}$, and $\phi K^{*}$.
}
\label{fig:Diagram_phiKst}
\end{center}
\end{figure}

The decays $B\to{\phi K^{*}}$ are expected
to proceed through pure penguin diagrams ($b\to s$ loops) as
illustrated in Fig.~\ref{fig:Diagram_phiKst}. 
Similarly, the decays $B\to{\rho K^{*}}$ are expected to be
dominated by $b\to s$ penguin transitions with additional contributions 
from Cabibbo-suppressed tree-level $b\to u$ transitions, while the decays 
$B\to{\rho\rho}$ proceed primarily through Cabibbo-favored
tree-level $b\to u$ transitions and CKM-suppressed $b\to d$ penguins.

%%%%%%%%%%%%%%%%%%%%%%%%%%%%%%%%%%%%%%
%%%%%%%%%%%%%%%%%%%%%%%%%%%%%%%%%%%%%%
\vspace{-0.5cm}

The measurement of direct $\CP$ violation in pure penguin modes, 
such as $B\to\phi{K}^{*}$, is sensitive to non-standard-model 
physics \cite{newphys}. In the standard model, direct $\CP$ violation 
could arise due to the difference between the 
$b\to u$ tree and $b\to s$ ($b\to d$) penguin amplitude weak 
phases~\cite{Bander}, which is $\gamma$ ($\alpha$) in the 
case of the decays $B\rightarrow\rho K^*$ ($B\rightarrow\rho\rho$).
Direct $\CP$ violation would be observed as an asymmetry of $B$ decay rates:
%%%%%%%%%%%%%%%%%%
\begin{eqnarray}
{\cal A}_{\CP}\equiv\frac{\Gamma(\Bbar\rightarrow\bar f)-\Gamma(B\rightarrow f)}
              {\Gamma(\Bbar\rightarrow\bar f)+\Gamma(B\rightarrow f)} \ .
\label{eq:acpdecay}
\end{eqnarray}
%%%%%%%%%%%%%%%%%%
However, large uncertainties in the strong phases,
which can be calculated by certain models,
weakens the quantitative relationship to the weak phases.

The time-dependent asymmetries in $B$ decays to $\CP$
eigenstates would provide important tests of the standard 
model~\cite{BCP}.
Comparison of the value of $\sin 2\beta$ obtained from 
$\phi K^{*0}({\rightarrow K^0_{S}\pi^0})$ with that from 
charmonium modes can probe for new physics. 
Time-dependent measurements in $B\to{\rho\rho}$ modes
combined with isospin relations among the decay amplitudes
for these modes would provide a measurement of $\alpha$.
Angular analysis is important for time-dependent asymmetries 
because of the mixture of $\CP$-odd and $\CP$-even components, 
and for the isospin analysis of $B\to{\rho\rho}$ modes.

%%%%%%%%%%%%%%%%%%%%%%%%%%%%%%%%%%%%%%%%%%%%%%%%%%%%%%%%%%%%%%%%%%%%%%
\section{THE \babar\ DETECTOR AND DATASET}
\label{sec:Data}

In this analysis we use
the data collected with the \babar\ detector~\cite{babar}
at the \pep2 asymmetric-energy $e^+e^-$ collider~\cite{pep}
located at the Stanford Linear Accelerator Center.
The results presented in this paper are based on data taken
in the 1999--2002 run comprising an integrated luminosity 
of 81.9~fb$^{-1}$, corresponding to 88.9 million 
$\BB$ pairs, at the $\FourS$ resonance
(on-resonance) and 9.6~fb$^{-1}$ approximately 
40~MeV below this energy (off-resonance). 
The $\FourS$ resonance occurs at the $e^+e^-$ 
center-of-mass (c.m.) energy, $\sqrt{s}$, of 10.58 GeV.
The asymmetric beam configuration in the laboratory frame provides 
a boost to the $\FourS$ increasing the momentum range of the 
$B$ meson decay products up to about 4~GeV/$c$.

Charged-particle momenta are measured in a tracking system 
that is a combination of a silicon vertex tracker (SVT) consisting 
of five double-sided detectors and a 40-layer central drift chamber 
(DCH), both operating in a 1.5-T solenoidal magnetic field. 
\babar\ achieves an impact parameter resolution
of about 40~$\mu$m for the high momentum charged particles
in the $B$ decay final states, allowing the precise determination 
of decay vertices.
The tracking system covers 92\% of the solid angle in the c.m. frame.
The average track-finding efficiency is approximately 98\% for 
momenta above 0.2~GeV/$c$ when the angle between track and 
the beam axis is greater than 500~mrad.

Charged particle identification is provided by the average 
energy loss ($dE/dx$) in the tracking devices (SVT and DCH) and
by an internally reflecting ring imaging Cherenkov detector 
(DIRC) covering the central region. A $K$--$\pi$ separation of 
better than four standard deviations ($\sigma$) is achieved for 
momenta below 3~GeV/$c$, decreasing to 2.5$\sigma$ at the highest 
momenta in the $B$ decay final states.
Photons are detected by a CsI(Tl) electromagnetic calorimeter (EMC), which
provides excellent angular and energy resolution with high efficiency for 
energies above 30~MeV. For a 1~GeV photon 
the energy and angular resolutions are $3\%$ and 4 mrad, respectively.
Electrons are identified by the tracking system and the EMC.

%%%%%%%%%%%%%%%%%%%%%%%%%%%%%%%%%%%%%%%%%%%%%%%%%%%%%%%%%%%%%%%%%%%%%%
\section{EVENT SELECTION}
\label{sec:Selection}

Hadronic events are selected based on track multiplicity and 
event topology. We fully reconstruct $B$ meson candidates 
from their charged and neutral decay products including
the intermediate states $\phi\to K^+K^-$, 
$K^{*0}\to K^+\pi^-$, $K^{*0}\to K^0\pi^0$,
$K^{*+}\to K^+\pi^0$, $K^{*+}\to K^0\pi^+$, 
$\rho^0\to \pi^+\pi^-$, $\rho^+\to \pi^+\pi^0$, with
$\pi^0\rightarrow \gamma\gamma$ and
$K^0\rightarrow K^0_S\rightarrow\pi^+\pi^-$ \cite{chgconj}.
Candidate charged tracks are required to originate 
from the interaction point, and to have at least 12 DCH hits 
and a minimum transverse momentum of 0.1~GeV/$c$. 
Looser criteria are applied to tracks forming $K^0_S$ candidates
to allow for displaced decay vertices.
The $K^0_S$ candidates are required to satisfy 
$|m_{\pi^+\pi^-} - m_{K^0}|<$ 12~MeV/$c^2$  
with the cosine of the angle between their reconstructed flight and
momentum directions greater than 0.995 and the measured proper 
decay time greater than five times its uncertainty.
Kaon tracks are distinguished from pion and proton tracks via a
likelihood ratio that includes, for momenta below 0.7~GeV/$c$, 
$dE/dx$ information from the SVT and DCH, and, for higher
momenta, the Cherenkov angle and number of photons
as measured by the DIRC. 

We reconstruct $\pi^0$ mesons from pairs of photons, each with 
a minimum energy 30 MeV. The typical width of the reconstructed 
$\pi^0$ mass is 7~MeV/$c^2$. A $\pm$15~MeV/$c^2$ interval is applied 
to select $\pi^0$ candidates. We select $\phi$, $K^*$, and $\rho$ 
candidates with the following requirements on the invariant 
masses of their final states: 
$0.99 < m_{K^+K^-} < 1.05$, $0.75 < m_{K\pi} < 1.05$, and
$0.52 < m_{\pi\pi} <1.00$ (all in GeV/$c^2$).
The helicity angle $\theta_{\rm x}$ of a $\phi$, $K^*$, or $\rho$
(${\rm x}=1, 2$ for the two resonances in the $B$ decay)
is defined as the angle between the direction of one
of the two daughters ($K^+$, $K$, or $\pi^+$ respectively) 
and the parent $B$ direction in 
the resonance rest frame. To suppress combinatorial 
background we restrict the $K^{*}\rightarrow K\pi^0$ and 
$\rho^{+}\rightarrow \pi^+\pi^0$ helicity angles 
($\cos\theta_{\rm x} > -0.5$). 
This effectively requires the $\pi^0$ momentum 
to be larger than 0.35~GeV/$c$.

We identify $B$ meson candidates kinematically
using two nearly independent variables \cite{babar},
the beam energy-substituted mass $m_{\rm{ES}} =$ 
$[{ (s/2 + \mathbf{p}_i \cdot \mathbf{p}_B)^2 / E_i^2 - 
\mathbf{p}_B^{\,2} }]^{1/2}$ and the energy difference
$\Delta E = (E_i E_B - \mathbf{p}_i 
\cdot \mathbf{p}_B - s/2)/\sqrt{s}$,
where $(E_i,\mathbf{p}_i)$ is the initial state four-momentum
obtained from the beam momenta, and $(E_B,\mathbf{p}_B)$
is the four-momentum of the reconstructed $B$ candidate.
For signal events $m_{\rm{ES}}$ peaks at the $B$ mass 
and $\Delta E$ at zero. 
Our initial selection requires $m_{\rm{ES}}>5.2$ GeV/$c^2$
and $|\Delta E|<0.2$~GeV.

Charmless hadronic modes suffer from a large background due to 
random combinations of tracks produced in quark-antiquark  
continuum events ($e^+e^-\to\qqbar$, $q = u, d, s, c$).  
Background events from the continuum are distinguished 
by a jet-like structure as opposed to the more spherical 
topology of $\BB$ pairs produced in $\FourS$ events.
To reject continuum background 
we require $|\cos\theta_T| < 0.8$, where $\theta_T$ 
is the angle between the thrust axis of the $B$ candidate 
and that of the rest of the tracks and neutral clusters in
the event, calculated in the c.m. frame. The distribution of
the $\cos{\theta_T}$ variable is sharply peaked near $\pm1$ for 
combinations drawn from jet-like $\qqbar$ pairs, and nearly
uniform for the isotropic $B$ meson decays.
We also construct a Fisher discriminant that combines eleven 
variables~\cite{CLEO-fisher}: the polar angles of the $B$
momentum vector and the $B$-candidate thrust axis with respect 
to the beam axis in the $\FourS$ frame,
and the scalar sum of the c.m. momenta of charged particles
and photons (excluding particles from the $B$ candidate)
entering nine coaxial angular intervals of 10$^\circ$
around the $B$-candidate thrust axis.

Monte Carlo (MC) simulation \cite{geant} demonstrates that 
contamination from other $B$ decays is negligible for the 
modes with narrow $\phi$ resonance and is relatively small 
for other charmless $B$ decay modes. We achieve further 
suppression of $B$ decay background by removing all signal
candidates that have decay products consistent with
$D\rightarrow K\pi, K\pi\pi$ decays. The remaining small 
background coming from $B$ decays is accounted for in the fit. 
It is found that background subtraction is necessary
only in the $B\to\rho K^*$ analysis, where there is a small 
probability of charmed $B$ decays being reconstructed as signal.
In this analysis we assume negligible contribution of other
partial waves in our final states selected within vector 
resonance mass windows. 

%%%%%%%%%%%%%%%%%%%%%%%%%%%%%%%%%%%%%%%%%%%%%%%%%%%%%%%%%%%%%%%%%%%%%%
\section{ANALYSIS METHOD}
\label{sec:Fit}

We use an unbinned extended maximum likelihood (ML) fit to extract
signal yields, charge asymmetries, and angular polarizations
simultaneously. We define the likelihood for each event candidate:
%%%%%%%%%%%%%%%%%%%%%%%
\begin{equation}
{\cal L}_i = \sum_{j=1}^{3}\sum_{k=1}^{2}  n_{jk}\, 
{\cal P}_{jk}(\vec{x}_{i};\vec{\alpha}) ,
\label{eq:likelev}
\end{equation}
%%%%%%%%%%%%%%%%%%%%%%%
where ${\cal P}_{jk}(\vec{x}_{i};\vec{\alpha})$ is the probability 
density function (p.d.f.) for measured variables $\vec{x}_{i}$ of a
candidate $i$ in category $j$ and flavor state $k$, and $n_{jk}$ are 
the yields to be extracted from the fit.
There are three categories: signal ($j=1$), continuum~$\qqbar$
($j=2$), and $\BB$ combinatorial background ($j=3$).
The p.d.f.'s are non-zero only for the correct final 
state flavor ($k = 1$ for $\Bbar\rightarrow\bar f$ and 
$k = 2$ for $B\rightarrow f$).
The $B$ flavor is determined by its charge, except 
for the $\phi K^{*0}$ final state where 
the flavor is determined from the charge of the kaon 
from the $K^{*0}\to K^+\pi^-$ decay,
and the flavor is not defined for the
$K^{*0}\to K^0\pi^0$ decay.
The fixed parameters $\vec{\alpha}$ define the expected 
distributions of measured variables in each category and 
flavor state.
We rewrite the event yields $n_{jk}$ in each category in terms of 
the asymmetry ${\cal A}_j$ and the total event yield $n_{j}$:
$n_{j1} = n_{j}\times(1 + {\cal A}_j)/2$ and
$n_{j2} = n_{j}\times(1 - {\cal A}_j)/2$.
This definition is consistent with Eq.~\ref{eq:acpdecay}.

The fit input variables $\vec{x}_{i}$ are $\Delta E$, 
$m_{\rm{ES}}$, Fisher discriminant, invariant masses 
of the $K^*$ and $\phi$ (or $\rho$) resonances, and 
the $K^*$ and $\phi$ (or $\rho$) helicity angles
$\theta_{\rm x}$ (${\rm x}=1, 2$).
The correlations among the fit input variables 
in the data and signal MC are found to be small 
(typically less than $5\%$). The p.d.f. 
${\cal P}_{jk}(\vec{x}_{i};\vec{\alpha})$ 
for a given candidate $i$ is the product of the p.d.f.'s
for each of the variables, except for the helicity angles. 
We take into account the angular correlations
in the signal and the detector acceptance effects 
in the helicity angle p.d.f. parameterization. 
Due to the limited statistics in our analysis 
we adopt a simplified angular analysis technique 
where we integrate over the angle between the decay planes 
of the two vector-particle decays, leaving a p.d.f. that depends 
only on the two helicity angles. This distribution is sufficient 
to determine the longitudinal polarization fraction
$f_L\equiv{\Gamma_L}/{\Gamma}$.
The differential decay width is defined as:
%%%%%%%%%%%%%%%%%%%%%%%
\begin{eqnarray}
{1 \over \Gamma} \  {d^2\Gamma \over d\cos \theta_1 d\cos \theta_2} = 
{9 \over 4} \left \{ {1 \over 4} (1 - f_L)
\sin^2 \theta_1 \sin^2 \theta_2 + f_L \cos^2 \theta_1 \cos^2 \theta_2 \right\} \ .
\label{eq:helicityintegr}
\end{eqnarray}
%%%%%%%%%%%%%%%%%%%%%%%

We allow for multiple candidates in a given event by assigning 
candidates a weight of $1/N_i$, where $N_i$ is the number of 
candidates in the same event.
The average number of candidates per event is close to one 
(varying from 1.05 to 1.14 depending on the mode). 
The extended likelihood for a sample 
of $N_{\rm cand}$ candidates is 
%%%%%%%%%%%%%%%%%%%%%%%
\begin{equation}
{\cal L} = \exp\left(-\sum_{j=1}^{3} n_{j}\right)\, 
\prod_{i=1}^{N_{\rm cand}} 
\exp\left(\frac{\ln{\cal L}_i}{N_i}\right) .
\label{eq:likel}
\end{equation}
%%%%%%%%%%%%%%%%%%%%%%%

The event yields $n_j$, asymmetries ${\cal A}_j$, and
polarization $f_L$ are obtained by minimizing 
the quantity $\chi^2\equiv -2\ln{\cal L}$ using the 
minimization package {\tt MINUIT}~\cite{minuit}.
The dependence of $\chi^2$ on a fit parameter
$n_j$, ${\cal A}_j$, or $f_L$ is obtained with the other
fit parameters floating.
We quote statistical errors corresponding to unit
change in $\chi^2$.
When more than one $K^*$ decay channel is measured for the 
same $B$ decay, the channels are combined by adding their 
$\chi^2$ distributions for $n_j$, ${\cal A}_j$, or $f_L$.
The statistical significance of a signal is defined as the 
square root of the change in $\chi^2$ when constraining 
the number of signal events to zero in the likelihood fit;
it describes the probability for the background 
to fluctuate to the observed event yield.

The fixed parameters $\vec{\alpha}$ describing the p.d.f.'s 
for signal and background distributions are extracted 
from MC simulation, on-resonance $\Delta E$--$m_{\rm{ES}}$ 
sidebands, and off-resonance data. 
The MC resolutions are adjusted by comparisons of data and simulation 
in abundant calibration channels with similar kinematics and topology,
such as $B\rightarrow D\pi, D\rho$ with $D\rightarrow K\pi\pi, K\pi$.
The simulation reproduces the event-shape variable distributions
found in data.

To describe the signal distributions, we employ Gaussian functions 
for the parameterization of the p.d.f.'s for $\Delta E$ and
$m_{\rm{ES}}$ and a relativistic $P$-wave Breit-Wigner distribution
convoluted with a Gaussian resolution function for the resonance masses.
For the background we use low-degree polynomials or, in the case 
of $m_{\rm{ES}}$, an empirical phase-space function~\cite{argus}.
The background parameterizations for resonance masses also include 
a resonant component to account for resonance production in the 
continuum. The background helicity angle distribution shape is 
again separated into contributions from combinatorics and from 
real mesons, both fit by low-degree polynomials multiplied 
by an empirical function $1/(1+\exp((\theta_{\rm x}-\theta_0)/a))$
to account for the detector acceptance effects, where 
$\theta_0$ and $a$ are fixed parameters. 
For both the signal and background, the p.d.f. for the Fisher 
discriminant is represented by a Gaussian with different
widths above and below the mean.

%%%%%%%%%%%%%%%%%%%%%%%%%%%%%%%%%%%%%%%%%%%%%%%%%%%%%%%%%%%%%%%%%%%%%%
\section{PHYSICS RESULTS AND SYSTEMATIC UNCERTAINTIES}
\label{sec:Results}

\begin{figure}[htbp]
\begin{center}
\vspace{1cm} 

{\Large
(a) $B^0\rightarrow\phi K^{*0}$
~~~~~~~~~~~~~~~~~~~~~~~~~~~
(b) $B^+\rightarrow\phi K^{*+}$

\vspace{-10mm} }

\centerline{
\epsfig{figure=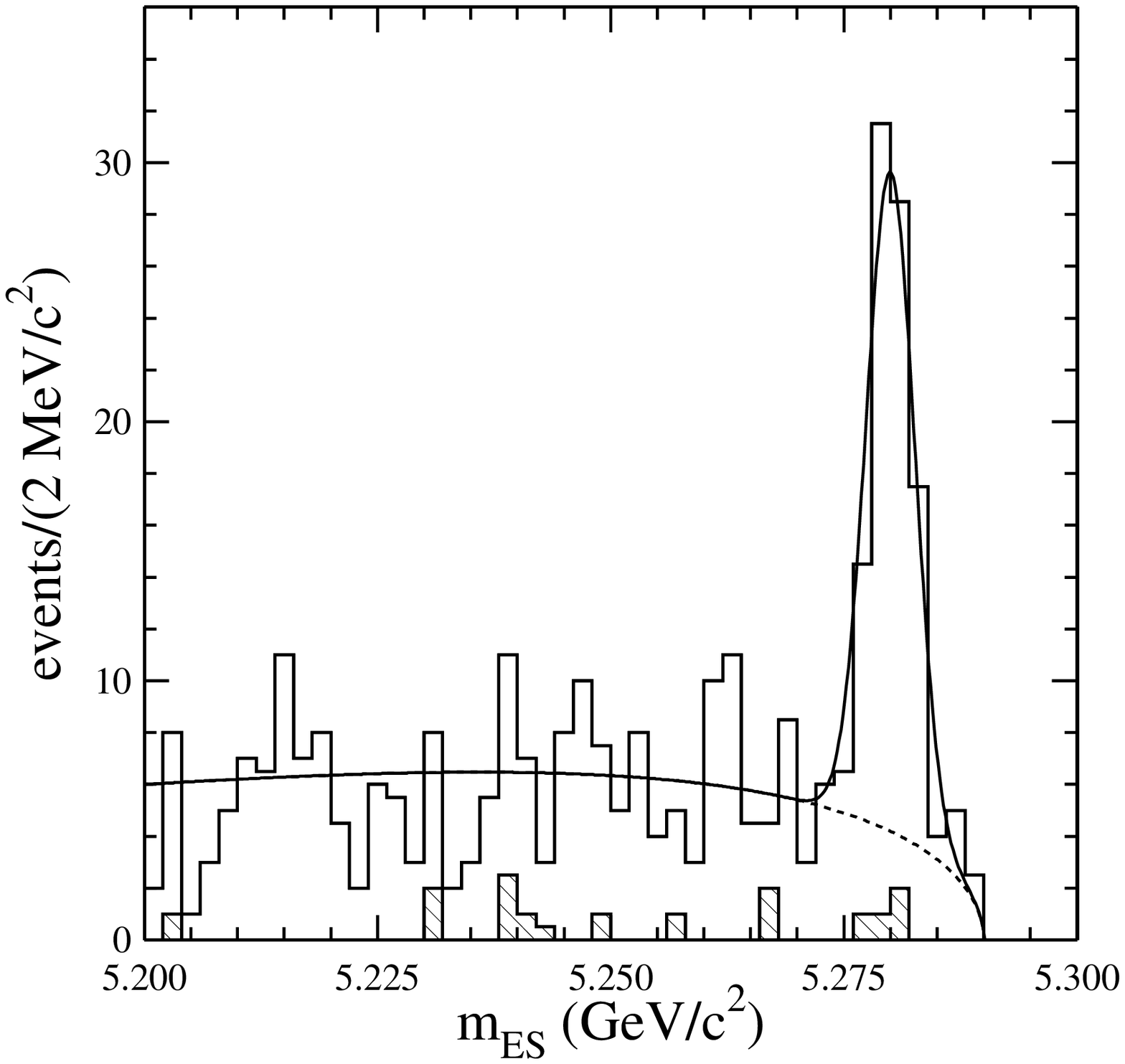,height=3.0in}
\epsfig{figure=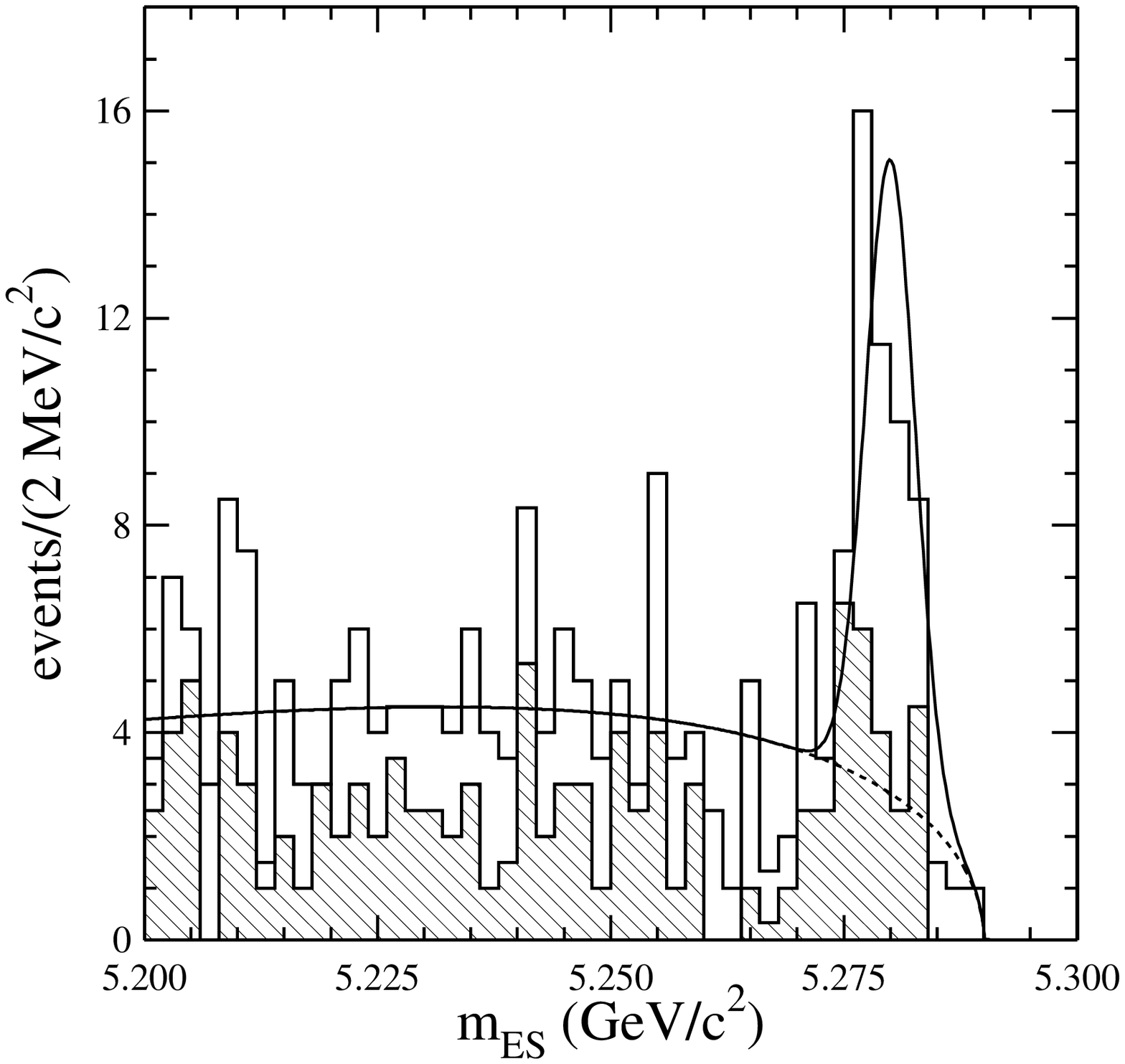,height=3.0in}
}
\vspace{1.0cm}

{\Large
(c) $B^+\rightarrow\rho^0 K^{*+}$
~~~~~~~~~~~~~~~~~~~~~~~~~~~~
(d) $B^+\rightarrow\rho^0 \rho^{+}$

\vspace{-10mm} }

\centerline{
\epsfig{figure=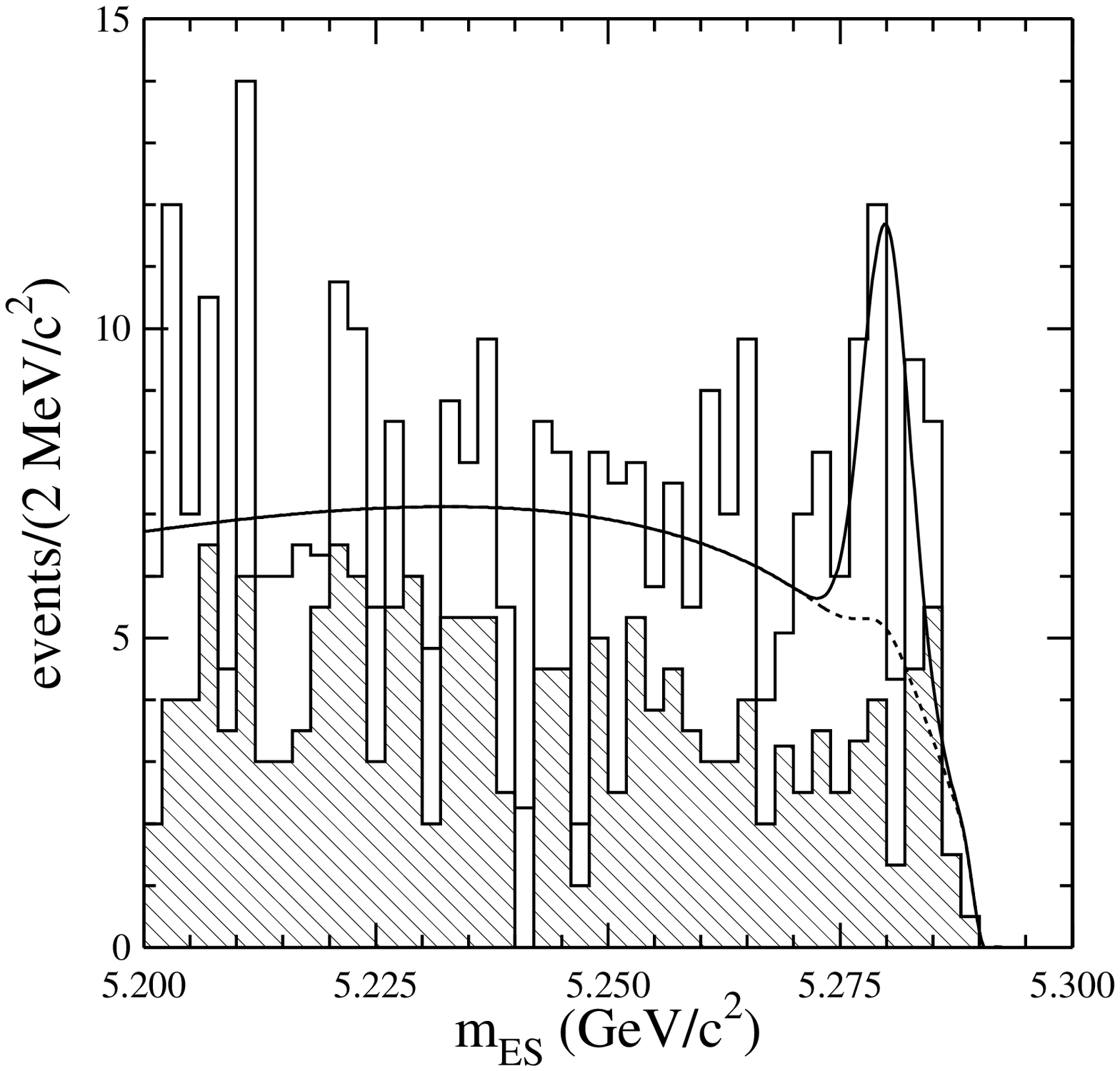,height=3.0in}
\epsfig{figure=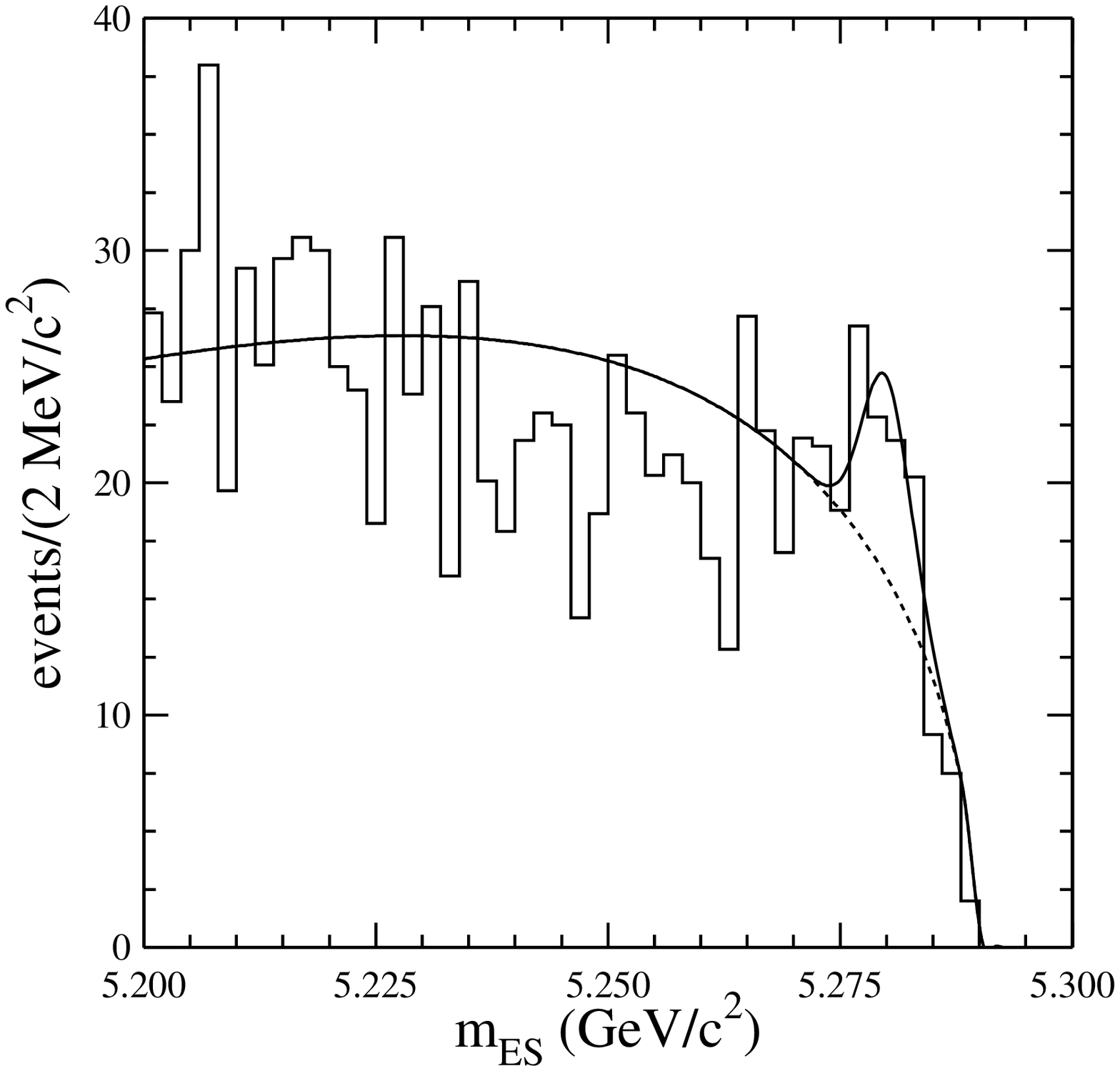,height=3.0in}
}
\vspace{0.5cm}
\caption{\sl 
Projections onto the variable $m_{\rm{ES}}$ for 
$B^0\rightarrow\phi K^{*0}$ (a), 
$B^+\rightarrow\phi K^{*+}$ (b),
$B^+\rightarrow\rho^0 K^{*+}$ (c), and 
$B^+\rightarrow\rho^0 \rho^{+}$ (d) candidates
after a requirement on 
the signal-to-background probability ratio
${\cal P}_{\rm{sig}}/{\cal P}_{\rm{bkg}}$ with the 
p.d.f. for $m_{\rm{ES}}$ excluded.
The histograms show the data, which are the sum of the two
$K^{*}$ decay channels when appropriate, while the shaded 
area is $K^{*}\rightarrow{K\pi^0}$ channel alone.
The solid (dashed) line shows the signal-plus-background 
(background only) p.d.f. projection.
}
\label{fig:mbproj_phiKst}
\end{center}
\end{figure}

\begin{figure}[htbp]
\begin{center}
\vspace{1cm} 

{\Large
(a)~~~~~~~~~~~~~~~~~~~~~~~~~~~~~~~~~~~~~~~~~~~~~~~(b)~~~~~~~~~~~~~~~~~~
\vspace{-10mm} }

\centerline{
\epsfig{figure=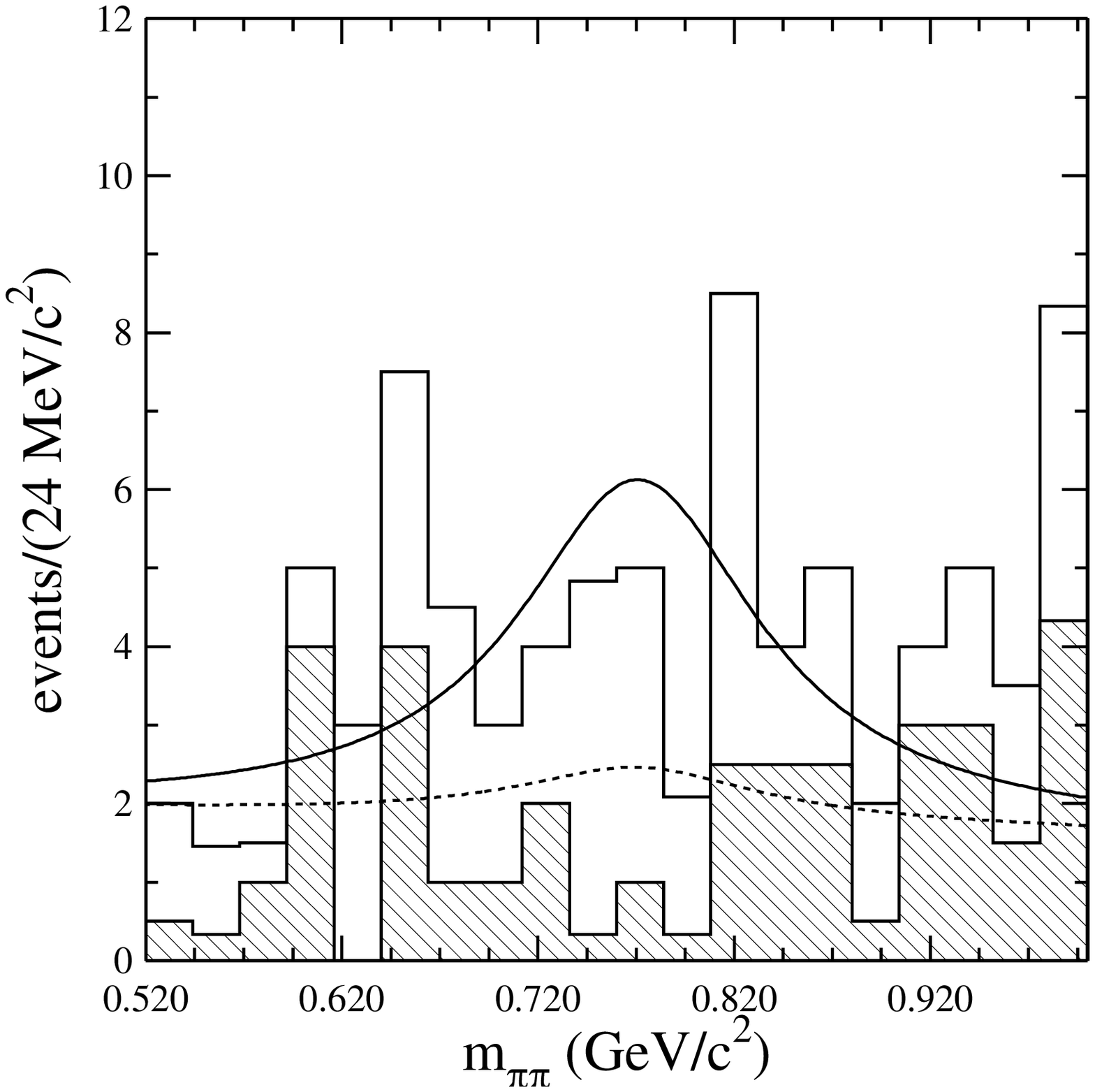,height=3.0in}
\epsfig{figure=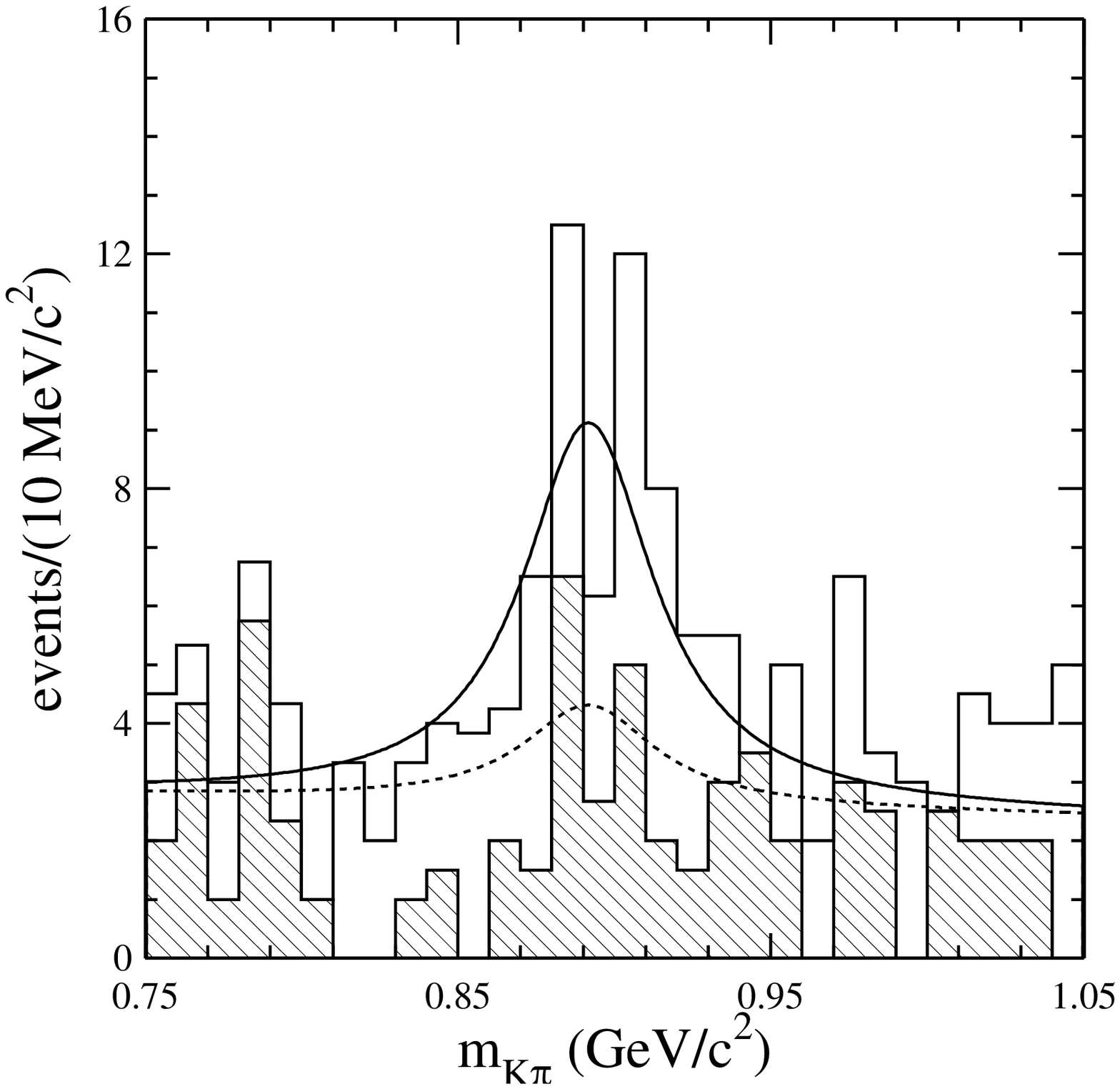,height=3.0in}
}
\vspace{1.0cm}

{\Large
(c)~~~~~~~~~~~~~~~~~~~~~~~~~~~~~~~~~~~~~~~~~~~~~~~(d)~~~~~~~~~~~~~~~~~~
\vspace{-10mm}} 

\centerline{
\epsfig{figure=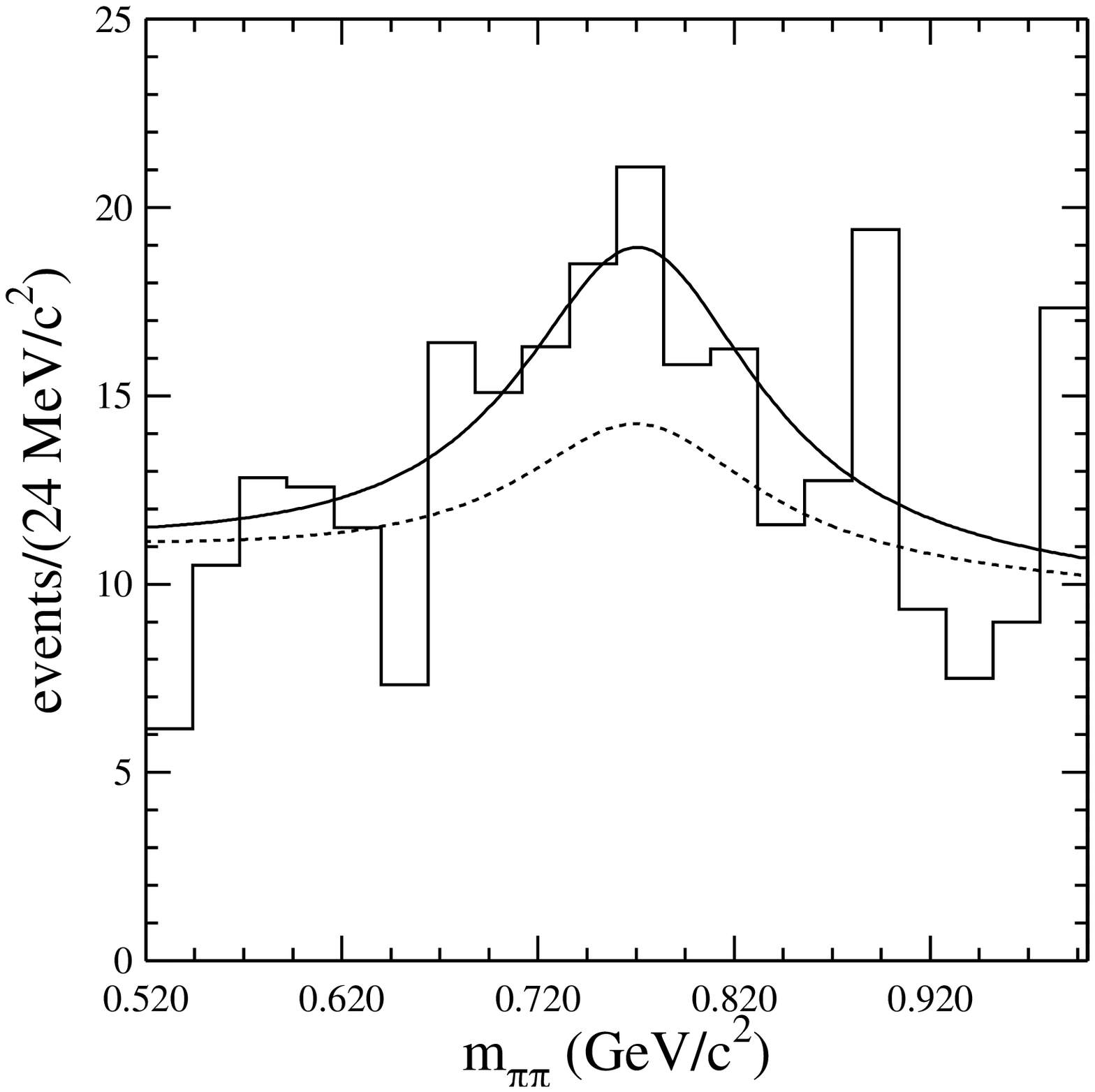,height=3.0in}
\epsfig{figure=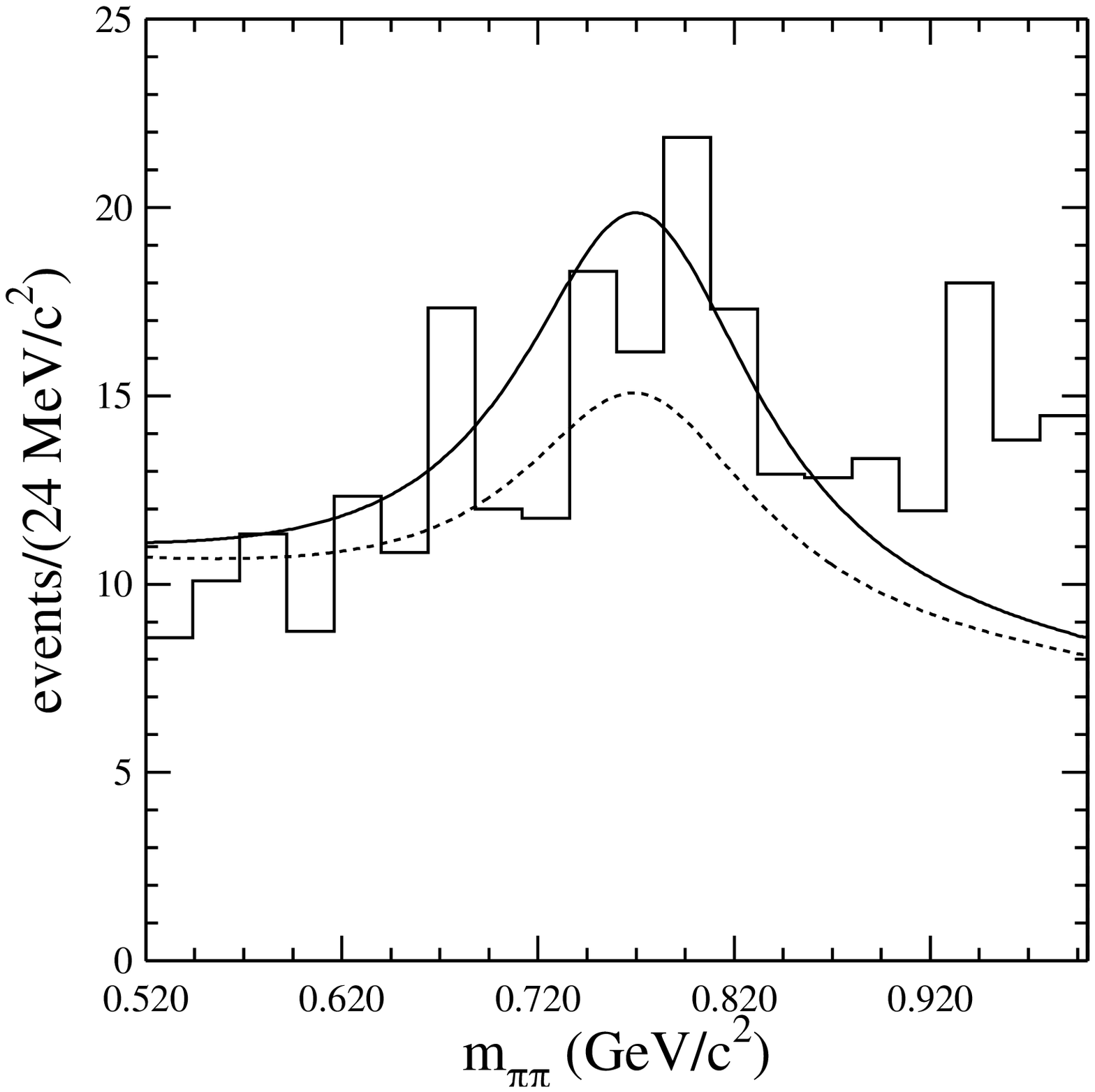,height=3.0in}
}
\vspace{0.5cm}
\caption{\sl 
Vector resonance invariant mass distributions
for $B^+\rightarrow\rho^0 K^{*+}$ candidates:
$m_{\pi^+\pi^-}$ (a) and $m_{K\pi}$ (b); and 
for $B^+\rightarrow\rho^0 \rho^{+}$ candidates:
$m_{\pi^+\pi^-}$ (c) and $m_{\pi^+\pi^0}$ (d),
after a requirement on 
the signal-to-background probability ratio
${\cal P}_{\rm{sig}}/{\cal P}_{\rm{bkg}}$ with the 
p.d.f. for the invariant mass excluded.
The histograms show the data and the shaded area 
corresponds to the $\rho^0 K^{*+}$ final state 
with $K^{*+}\rightarrow{K^+\pi^0}$ channel alone.
The solid (dashed) line shows the signal-plus-background 
(background only) p.d.f. projection.
}
\label{fig:mvproj_rhoKst}
\end{center}
\end{figure}

The results of our maximum likelihood fits are summarized in 
Table~\ref{tab:results}.
For the branching fractions we assume equal production rates of 
$\BzBzb$ and $\BpBm$.
We find significant signals in both $\phi K^*$ decay modes.
We measure the charge asymmetries and longitudinal 
polarizations in all $\phi K^*$ final states.
We also observe a significant yield of events in 
$\rho^0 K^{*+}$ (4.7$\sigma$)  
and $\rho^0\rho^+$ (4.4$\sigma$) final states.
The projections of the fit results are shown in 
Fig.~\ref{fig:mbproj_phiKst} and \ref{fig:mvproj_rhoKst},
where we plot only a subsample of events, enhancing the signal 
with a requirement on the ratio of the signal probability 
to background probability 
(${\cal P}_{\rm{sig}}$~and~${\cal P}_{\rm{bkg}}$ 
from Eq.~\ref{eq:likelev}).

We study the performance of the ML fit with the MC samples
where the signal events are taken from the complete MC simulation 
and the background is distributed according to the sideband 
parameterizations. 
%%%%%%%%%%%%%%%%%%%%%%%%%%%%%%%%%%%%%%
\begin{table}[h!]
\caption
{\sl Summary of results for the measured $B$ decay modes;
$\varepsilon$ denotes the reconstruction efficiency and
$\varepsilon_{\rm{tot}}$ 
the total efficiency including daughter branching fractions, both in percent;
$n_{\rm sig}$ is the fitted number of signal events, 
${\cal B}$ is the branching fraction, 
${\cal A}_{\CP}$ is the signal charge asymmetry, and 
$f_L$ is the longitudinal polarization.
The decay channels of $K^*$ are shown when more than one
final state is measured for the same $B$ decay mode.
All results include systematic errors, which are quoted 
following the statistical errors.
 }
\label{tab:results}
\scriptsize
\footnotesize
\begin{center}
\begin{tabular}{lcccccc}
\hline\hline
\vspace{-2.5mm}&&&\\
Mode & $\varepsilon$ & $\varepsilon_{\rm tot}$ 
 & $n_{\rm sig}$ 
 & ${\cal B}$ ($\times 10^{-6}$)& ${\cal A}_{\CP}$ & $f_L$ \cr
\vspace{-2.5mm}&&&\\
\hline
\vspace{-2.5mm}&&&\\
$\phi K^{*+}$ 
 & -- & 5.3 & -- & $12.1^{~+2.1}_{~-1.9}\pm 1.5$  
 & $+0.16\pm 0.17\pm 0.04$ & $0.46\pm 0.12\pm 0.05$ \cr
\vspace{-2.5mm}&&&\\
\hline
\vspace{-2.5mm}&&&\\
~~~$\rightarrow$${K^0\pi^+}$ 
 & 26.0 & 2.9 & $33.3^{+7.2}_{-6.4}\pm 1.2$  & $12.7^{~+2.8}_{~-2.5}\pm 1.2$  
 & $-0.02\pm 0.20\pm 0.03$ & $0.50^{~+0.14}_{~-0.15}\pm 0.04$ \cr
\vspace{-2.5mm}&&&\\
~~~$\rightarrow$${K^+\pi^0}$ 
 & 14.3 & 2.3 & $22.3^{+7.5}_{-6.5}\pm 3.2$  & $10.7^{~+3.6}_{~-3.1}\pm 1.9$  
 & $+0.63^{~+0.25}_{~-0.31}\pm 0.05$ & $0.40^{~+0.20}_{~-0.19}\pm 0.07$ \cr
\vspace{-2.5mm}&&&\\
\hline
\vspace{-2.5mm}&&&\\
$\phi K^{*0}$ 
 & -- & 10.2 & -- & $11.1^{~+1.3}_{~-1.2}\pm 1.1$  
 & $+0.04\pm 0.12\pm 0.02$ & $0.65\pm 0.07\pm 0.04$ \cr
\vspace{-2.5mm}&&&\\
\hline
\vspace{-2.5mm}&&&\\
~~~$\rightarrow$${K^+\pi^-}$ 
 & 29.7 & 9.7 & $101^{+12}_{-11}\pm 3$  & $11.7\pm 1.4\pm 1.1$  
 & $+0.04\pm 0.12\pm 0.02$ & $0.64\pm 0.07\pm 0.03$ \cr
\vspace{-2.5mm}&&&\\
~~~$\rightarrow$${K^0\pi^0}$ 
 & 11.4 & 0.6 & $2.0^{+3.4}_{-1.3}\pm 0.6$ & $~3.5^{~+6.1}_{~-2.3}\pm 1.1$  
 & -- & $1.00^{~+0.00}_{~-0.66}\pm 0.25$ \cr
\vspace{-2.5mm}&&&\\
\hline
\vspace{-2.5mm}&&&\\
$\rho^0 K^{*+}$ 
 & -- & 8.4 & -- & $7.7^{+2.1}_{-2.0}\pm 1.4$ & -- & -- \cr  
\vspace{-2.5mm}&&&\\
\hline
\vspace{-2.5mm}&&&\\
~~~$\rightarrow$${K^0\pi^+}$ 
 & 21.0 & 4.8 & $44.4^{+12.5}_{-11.4}\pm 3.4$  & $10.4^{+2.9}_{-2.7}\pm 1.7$ & -- & -- \cr
\vspace{-2.5mm}&&&\\
~~~$\rightarrow$${K^+\pi^0}$ 
 & 10.9 & 3.6 & $9.1^{+11.1}_{-9.6}\pm 5.2$  & $2.9^{+3.5}_{-3.0}\pm 1.8$ & -- & -- \cr
\vspace{-2.5mm}&&&\\
\hline
\vspace{-2.5mm}&&&\\
$\rho^0\rho^+$ 
 & 11.3 & 11.1 & $97.5^{+26.1}_{-24.3}\pm 12.1$  & $9.9^{+2.6}_{-2.5}\pm 2.5$ & -- & -- \cr
\vspace{-2.5mm}&&&\\
\hline\hline
\end{tabular}
\end{center}
\end{table}
%%%%%%%%%%%%%%%%%%%%%%%%%%%%%%%%%%%%%%
%%%%%%%%%%%%%%%%%%%%%%%%%%%%%%%%%%%%%%
%%%%%%%%%%%%%%%%%%%%%%%%%%%%%%%%%%%%%%
%\vspace{-0.5cm}
%%%%%%%%%%%%%%%%%%%%%%%%%%%%%%%%%%%%%%
%%%%%%%%%%%%%%%%%%%%%%%%%%%%%%%%%%%%%%
%\vspace{-0.5cm}
The input values for the number of signal events, charge 
asymmetry, and decay polarization are well reproduced, and any 
small discrepancy is accounted for in the systematic errors.

Systematic uncertainties in the ML fit originate from assumptions 
about the signal and background distributions. Uncertainties in
the p.d.f. parameters arise from the limited statistics in the
background sideband data and signal control samples.
We vary the p.d.f. parameters within their respective uncertainties,
and derive the associated systematic errors.
The signals remain statistically significant under
these variations. Additional systematic errors in the number
of signal events originate from uncertainty in the charm
background subtraction in the $\rho K^*$ channels where we 
take the uncertainty to be 100$\%$ of the subtracted values 
(2 and 5 events in $K^{*+}\rightarrow K^0\pi^+$ and 
$K^{*+}\rightarrow K^+\pi^0$ channels respectively).

The dominant systematic errors in the efficiency are track 
finding (0.8\% per track), particle identification (2\% per track), 
and $K^0_S$ and $\pi^0$ reconstruction (5\% each). 
Other minor systematic effects from event selection criteria, 
daughter branching fractions~\cite{pdg}, MC statistics, 
and $B$ meson counting sum to less than 4\%.
The efficiency in the ML fit to signal 
samples can be less than 100\% because of fake combinations
passing the selection criteria, and we account for this with a 
systematic uncertainty (3--12\%).
This effect is larger in the final states with $\rho$
resonances because of the broader distributions.
Efficiency uncertainties affect the value of the branching 
fraction, but not its significance.

In the search for $\rho K^{*}$ and $\rho\rho$ final states
we fit only for event yields and
exclude angular and $B$-flavor information from the analysis.
The reconstruction efficiency depends on the decay 
polarization. We average reconstruction efficiencies for the 
$100\%$ transverse and $100\%$ longitudinal angular 
polarizations for each decay channel and assign 
the systematic errors as the root-mean-square of the 
uniform efficiency distributions between the two extreme cases.
The resulting systematic errors are 
9\% in $\rho^0 K^{*+}(\rightarrow{K^0\pi^+})$, 
19\% in $\rho^0 K^{*+}(\rightarrow{K^+\pi^0})$, 
and 18\% in $\rho^0\rho^+$ decay channels. 
For the $\phi K^*$ final states we calculate the efficiencies
assuming an average polarization of ($0.60\pm 0.06\pm 0.05$)
and assign a systematic error corresponding to the total 
polarization measurement error.

In the polarization measurements, we again include uncertainties 
from p.d.f. variations where we account for possible bias 
in the detector acceptance and background parameterizations.
The biases from the finite resolution in helicity angle measurement 
and dilution due to the presence of the fake combinations are
studied with MC simulation and are accounted for with conservative
systematic error of 0.03.

We find the charge asymmetry of the track reconstruction efficiency 
to be consistent with zero within an uncertainty of less than 0.01 
for a wide range of momenta~\cite{prdacp}. 
Taking into account particle identification 
requirements similar to the ones applied to the $K^*$ daughters, 
the asymmetry is consistent with zero with an uncertainty of 0.02.
We also find a negligible effect on the measured asymmetry 
from any possible bias in track-momentum measurements 
studied in $e^+e^-\to\mu^+\mu^-$ and cosmic ray events. 
The asymmetry measurement in the $B^0\to\phi K^{*0}$
decay mode is corrected by the inverse dilution factor $1/(1-2w)$,
where $w$, the fraction of doubly misidentified $K\pi$
combinations originating from $K^{*0}$, is less than $0.01$.

%%%%%%%%%%%%%%%%%%%%%%%%%%%%%%%%%%%%%%%%%%%%%%%%%%%%%%%%%%%%%%%%%%%%%%
\section{SUMMARY}
\label{sec:Conclusions}

We have measured branching fractions, longitudinal polarizations,
and charge asymmetries with the decays $B^0\to\phi K^{*0}$ and 
$B^+\to\phi K^{*+}$. 
Because the final states contain three strange quarks or antiquarks, 
in the standard model they are necessarily due to penguin diagrams.
This makes them particularly susceptible to non-standard-model 
contributions. We observe the decays $B^+\to\rho^0 K^{*+}$ and 
$B^+\to\rho^0\rho^{+}$ and report the corresponding branching
fractions. The $B^+\to\rho^0 K^{*+}$ decay process is of interest 
for direct $\CP$ measurements due to possible large penguin-tree 
interference, while the $B\to\rho\rho$ decays have potential 
for the measurement of the weak phase~$\alpha$.

The $B\to\phi K^{*}$ branching fractions are in agreement with 
our earlier less precise measurements~\cite{prlphik}. 
Our charge asymmetry results rule out a significant part of the 
physical ${\cal A}_{\CP}$ region, allowing for constraints 
on new physics models~\cite{newphys}, but are not yet of 
sufficient precision to allow precise comparison with 
standard model predictions~\cite{smphys}. 
We have performed the angular analysis in the penguin-dominated 
rare $B$ decays and measure a relatively large longitudinal 
polarization in the decay amplitude, as predicted~\cite{bvv}. 
Our measurement of $B^+\to\rho^0\rho^{+}$ 
branching fraction is significantly 
lower than the central value measured by Belle~\cite{belle}
even taking into account polarization uncertainty.
Our results are preliminary and use increased statistics compared
to the earlier $\babar$ measurements of the $B\to\phi K^{*}$ 
branching fractions and charge asymmetries~\cite{prlphik, prdacp}.

%%%%%%%%%%%%%%%%%%%%%%%%%%%%%%%%%%%%%%%%%%%%%%%%%%%%%%%%%%%%%%%%%%%%%%
\section{ACKNOWLEDGMENTS}
\label{sec:Acknowledgments}

We are grateful for the 
extraordinary contributions of our \pep2\ colleagues in
achieving the excellent luminosity and machine conditions
that have made this work possible.
The success of this project also relies critically on the 
expertise and dedication of the computing organizations that 
support \babar.
The collaborating institutions wish to thank 
SLAC for its support and the kind hospitality extended to them. 
This work is supported by the
US Department of Energy
and National Science Foundation, the
Natural Sciences and Engineering Research Council (Canada),
Institute of High Energy Physics (China), the
Commissariat \`a l'Energie Atomique and
Institut National de Physique Nucl\'eaire et de Physique des Particules
(France), the
Bundesministerium f\"ur Bildung und Forschung and
Deutsche Forschungsgemeinschaft
(Germany), the
Istituto Nazionale di Fisica Nucleare (Italy),
the Foundation for Fundamental Research on Matter (The Netherlands),
the Research Council of Norway, the
Ministry of Science and Technology of the Russian Federation, and the
Particle Physics and Astronomy Research Council (United Kingdom). 
Individuals have received support from 
the A. P. Sloan Foundation, 
the Research Corporation,
and the Alexander von Humboldt Foundation.

%%%%%%%%%%%%%%%%%%%%%%%%%%%%%%%%%%%%%%%%%%%%%%%%%%%%%%%%%%%%%%%%%%%%%%
\renewcommand{\baselinestretch}{1}

%%%%%%%%%%%%%%%%%%%%%%%%%%%%%%%%%%%%%%%%%%%%%%%%%%%%%%%%%%%%%%%%%%%%%%

\begin{thebibliography}{99}

\bibitem{cleo}
\label{ref:cleo}
The CLEO Collaboration, S.~Chen {\it et al.}, 
Phys.\ Rev.\ Lett. {\bf 85}. 525 (2000). 

\bibitem{prlphik}
The $\babar$ Collaboration, B.~Aubert {\it et al.},
Phys.\ Rev.\ Lett. {\bf 87}, 151801 (2001).

\bibitem{prdacp}
The $\babar$ Collaboration, B.~Aubert {\it et al.},
Phys.\ Rev.\ D {\bf 65}, 051101 (2002).

\bibitem{cleorhokst}
The CLEO Collaboration, R. Godang {\it et al.},
Phys.\ Rev.\ Lett. {\bf 88}, 021802 (2002).

\bibitem{belle}
The Belle Collaboration, K.~Abe {\it et al.}, BELLE-CONF-0255.

\bibitem{Kobayashi}
\label{ref:Kobayashi}
M.~Kobayashi and T. Maskawa, Prog. Theor. Phys. {\bf 49}, 652 (1973).

\bibitem{bvv}
G. Kramer, W.F. Palmer, Phys.\ Rev. D{\bf 45}, 193 (1992);
R.~Aleksan {\it et al.}, 
Phys.\ Lett.\ B {\bf 356}, 95 (1995);
C.-H. Chen, Y.-Y. Keum, H.-n. Li,
Phys.\ Rev.\ D {\bf 66}, 054013 (2002).

\bibitem{newphys}
\label{ref:newphys}
I.~Hinchliffe, N.~Kersting, Phys. Rev. D {\bf 63}, 015003 (2001).
%hep-ph/0003090

\bibitem{Bander}
\label{ref:Bander}
M.~Bander, D.~Silverman, and A.~Soni, Phys.\ Rev.\ Lett. {\bf 43}, 242 (1979).

\bibitem{BCP}
A.B.~Carter and A.I.~Sanda, Phys.\ Rev. {\bf D23}, 1567 (1981); I.I.~Bigi
and A.I.~Sanda, Nucl.\ Phys. {\bf B193}, 85 (1981).

\bibitem{babar}
The \babar\ Collaboration, B.~Aubert {\it et al.},
Nucl.\ Instrum.\ Methods {\bf A479}, 1 (2002).

\bibitem{pep} 
PEP-II Conceptual Design Report, SLAC-R-418 (1993).

\bibitem{chgconj} 
Inclusion of the charge conjugate states is implied.

\bibitem{CLEO-fisher}
The CLEO Collaboration,
D.M.~Asner {\it et al.}, 
Phys.\ Rev.\ D {\bf 53}, 1039 (1996).

\bibitem{geant}
The \babar\ detector Monte Carlo 
simulation is based on GEANT:
R.~Brun {\it et al.}, CERN DD/EE/84-1.

\bibitem{minuit}
F.~James,
CERN Program Library, D506.

\bibitem{argus}
The ARGUS Collaboration, H.~Albrecht {\it et al.}, Phys.\ Lett.\ B {\bf 241}, 278 (1990).

\bibitem{pdg} The Particle Data Group,
K. Hagiwara {\it et al.}, Phys. Rev. D {\bf 66}, 010001 (2002).

\bibitem{smphys}
\label{ref:smphys}
G.~Kramer, W.F.~Palmer, and H.~Simma,
Nucl.\ Phys. B {\bf 428}, 77 (1994);
A.~Ali, G.~Kramer, and C.-D.~L\"{u},
Phys.\ Rev. D {\bf 59}, 014005 (1998).

\end{thebibliography}
\end{document}